# A NON LINEAR MULTIFRACTAL STUDY TO ILLUSTRATE THE EVOLUTION OF TAGORE SONGS OVER A CENTURY

Shankha Sanyal*[1,2], Archi Banerjee[1,2], Tarit Guhathakurata[1],
Ranjan Sengupta[1] and Dipak Ghosh[1]

[1]Sir C.V. Raman Centre for Physics and Music, [2]Department of Physics
Jadavpur University, Kolkata: 700032
*ssanyal.sanyal2@gmail.com

## ABSTRACT:
*The works of Rabindranath Tagore have been sung by various artistes over generations spanning over almost 100 years. there are few songs which were popular in the early years and have been able to retain their popularity over the years while some others have faded away. In this study we look to find cues for the singing style of these songs which have kept them alive for all these years. For this we took 3 min clip of four Tagore songs which have been sung by five generation of artistes over 100 years and analyze them with the help of latest nonlinear techniques Multifractal Detrended Fluctuation Analysis (MFDFA). The multifractal spectral width is a manifestation of the inherent complexity of the signal and may prove to be an important parameter to identify the singing style of particular generation of singers and how this style varies over different generations. The results are discussed in detail.*

## INTRODUCTION:

The great visionary from Bengal, Rabindranath Tagore once said *"Whatever fate may be in store in the judgment of the future for my poems, my stories and my plays, I know for certain that the Bengali race must accept my songs, they must all sing my songs in every Bengali home, in the fields and by the rivers... I feel as if music wells up from within some unconscious depth of my mind, that is why it has certain completeness"* [1]. For Bengalis, their appeal—arousing from the combination of emotive strength and beauty described as surpassing even Tagore's poetry—was such that the *Modern Review*[2] observed that *"there is in Bengal no cultured home where Rabindranath's songs are not sung or at least attempted to be sung ... Even illiterate villagers sing his songs."* What is the secret formula which makes Tagore songs, which were composed more than 100 years back, survive and flourish in this age of computer generated music? There have been very few studies which try to scientifically analyze the evolution of Tagore songs over the years [3-5]. In one of these works [5], it was seen that the main cue for variation between singers of the same Tagore song lies on stress given to a particular syllable as well as the glide (*meend*) time between two notes. In this work, we make use of the latest state-of-the-art non linear techniques to quantitatively assess the cues which are phenomenal in the evolution of Tagore songs over the last century.

The fractal and multifractal aspects of different genres of music were analyzed by Bigrelle and Iost [6], where it was proposed that the use of fractal dimension measurements could benefit the discrimination of musical genres. Su and Wu [7] applied Hurst exponent and Fourier analysis in sequences of musical notes and noted that music shares similar fractal properties with the fractional Brownian motion. But, music signals may exhibit self similarity in different scales which may not be described by single scaling exponent as is found in Detrended Fluctuation Analysis (DFA) [8] technique. That is, the clustering pattern is not uniform over the whole system. Such a system is better characterized as 'multifractal'. A multifractal can be loosely thought of as an interwoven set constructed from sub-sets with different local fractal dimensions. Real world systems are mostly multifractal in nature. Music too, has non-uniform property in its movement. Since the multifractal technique analyzes the signal in different scales, it is able to decipher much accurately the amount of self similarity present in a signal. The spectrum in multifractal detrended fluctuation analysis (MFDFA) [9] is the measure of complexity or self-similarity present in the signal. This method is very useful in the analysis of various non-stationary time series and it also gives information regarding the multifractal scaling behaviour of non-stationary signals. We used this technique to characterize music samples collected for different Tagore songs sung over the century and classified them on the basis of their complexity values which is detrimental of the uniqueness of that particular song.

---
* Corresponding Author

Four widely popular songs of the maestro Rabindranath Tagore were taken for our analysis, which have been popular over the years, sung by singers of different generations. The songs are *Bohujuger Opar Hote, Maharaj Eki Saje, Krishnakoli Ami Tarei Boli* and *Amaro Porano Jaha Chay*. The rationale behind choosing these songs is that these few compositions have sustained through the ages and is still being sung by singers of this era, while many others have lost their popularity. The multifractal width (or the complexity values) was computed for each of the phrases and their variation was studied over the years to identify cues for evolution. The study reveals interesting results regarding the style of rendition of each individual performer or a few performers of the same generation. For e.g. while the singers of the first generation (when there were no formal musical notation of the songs) showed significantly high values of complexity throughout their rendition, singers of the latter generations showed a decrease in the complexity values. Another interesting observation is that, those singers who have been trained in *Vishwa Bharati* or under the patronage of the maestro himself show a distinct lineage or unique style of rendition clearly distinguishable from the others. Thus, with this study, we have tried to quantify and visualize how singers have rendered their unique style of singing the same song, so as to keep the popularity value intact but without losing the flavor of *Rabindrasangeet*. The change in the singing style of the same song over the years by different artistes is also manifested in this study.

## EXPERIMENTAL DETAILS:

Four widely popular songs of the maestro Rabindranath Tagore were taken for our analysis, which have been popular over the years, sung by singers of different generations. The different songs taken for our analysis have been listed in **Table 1** in a chronological order.

**Table 1:** Chronological list of Songs taken for our analysis

|  | Artist | Year of Recording |
|---|---|---|
| Bahujuger Opar Hote | Bimalbhusan Mukherjee | 1935 |
|  | Kanak Das | 1956 |
|  | Debabrata Biswas | 1973 |
|  | Mita Hoq | 1986 |
|  | Rezwana Chowdhury | 2002 |
| Maharaj Eki Saje | Amala Das | 1914 |
|  | Subinoy Roy | 1945 |
|  | Debabrata Biswas | 1963 |
|  | Mohan Singh | 1977 |
|  | Shahana Bajpayee | 2008 |
| Amaro Porano Jaha Chai | Dinendra Thakur | 1921 |
|  | Sagar Sen | 1946 |
|  | Chinmoy Chatterjee | 1963 |
|  | Suchitra Mitra | 1979 |
|  | Jayati Chakraborty | 2002 |
| Krishnakoli Ami Tarei Boli | Shantidev Ghosh | 1950 |
|  | Suchitra Mitra | 1976 |
|  | Hemanta Mukherjee | 1985 |
|  | Kamalini Mukherjee | 1999 |
|  | Sasha Ghoshal | 2010 |

The signals are digitized at the rate of 22050 samples/sec 16 bit format. 3 min clips of each song were taken, and similar phases were extracted (which were about 45 seconds duration) from the rendition of each generation for our analysis. The multifractal width (or the complexity values) was computed for each of the phrases and their variation was studied over the years to identify cues for evolution.

# METHOD OF ANALYSIS
## Method of multifractal analysis of sound signals

The time series data obtained from the sound signals are analyzed using MATLAB [8] and for each step an equivalent mathematical representation is given which is taken from the prescription of Kantelhardt et al [7].

The complete procedure is divided into the following steps:

*Step 1:* Converting the noise like structure of the signal into a random walk like signal. It can be represented as:

$$Y(i) = \sum (x_k - \bar{x}) \quad (1)$$

Where $\bar{x}$ is the mean value of the signal.

*Step 2:* The local RMS variation for any sample size $s$ is the function $F(s,v)$. This function can be written as follows:

$$F^2(s,v) = \frac{1}{s}\sum_{i=1}^{s}\{Y[(v-1)s+i] - y_v(i)\}^2$$

*Step 4:* The q-order overall RMS variation for various scale sizes can be obtained by the use of following equation

$$F_q(s) = \left\{\frac{1}{Ns}\sum_{v=1}^{Ns}[F^2(s,v)]^{\frac{q}{2}}\right\}^{\left(\frac{1}{q}\right)} \quad (2)$$

*Step 5:* The scaling behaviour of the fluctuation function is obtained by drawing the log-log plot of $F_q(s)$ vs. s for each value of q.

$$F_q(s) \sim s^{h(q)} \quad (3)$$

The h(q) is called the generalized Hurst exponent. The Hurst exponent is measure of self-similarity and correlation properties of time series produced by fractal. The presence or absence of long range correlation can be determined using Hurst exponent. A monofractal time series is characterized by unique h(q) for all values of q.

The generalized Hurst exponent h(q) of MFDFA is related to the classical scaling exponent $\tau(q)$ by the relation

$$\tau(q) = qh(q) - 1 \quad (4)$$

A monofractal series with long range correlation is characterized by linearly dependent q order exponent $\tau(q)$ with a single Hurst exponent H. Multifractal signal on the other hand, possess multiple Hurst exponent and in this case, $\tau(q)$ depends non-linearly on q [9].

The singularity spectrum f(α) is related to h(q) by

$$\alpha = h(q) + qh'(q)$$
$$f(\alpha) = q[\alpha - h(q)] + 1$$

Where α denoting the singularity strength and *f(α)*, the dimension of subset series that is characterized by α. The width of the multifractal spectrum essentially denotes the range of exponents. The spectra can be characterized quantitatively by fitting a quadratic function with the help of least square method [9] in the neighbourhood of maximum $\alpha_0$,

$$f(\alpha) = A(\alpha - \alpha_0)^2 + B(\alpha - \alpha_0) + C \quad (5)$$

Here C is an additive constant C = f($\alpha_0$) = 1and B is a measure of asymmetry of the spectrum. So obviously it is zero for a perfectly symmetric spectrum. We can obtain the width of the spectrum very easily by extrapolating the fitted quadratic curve to zero.

Width W is defined as,

$$W = \alpha_1 - \alpha_2 \quad (6)$$

with $f(\alpha_1) = f(\alpha_2) = 0$

The width of the spectrum gives a measure of the multifractality of the spectrum. Greater is the value of the width W greater will be the multifractality of the spectrum. For a monofractal time series, the width will be zero as h(q) is independent of q. The spectral width has been considered as a parameter to evaluate how a group of string instruments vary in their pattern of playing from another

**RESULTS AND DISCUSSION:**

Every musical composition/element can be considered as a nonlinear complex time series – the multifractal width (w) being a quantitative measure of its complexity. In other words, more w – more local fluctuations in temporal scale and thus this parameter is very sensitive to characterize and quantify a particular music signal to a level which is not possible with any other method. In a similar manner, small w implies less local fluctuations in temporal scale. Thus, similar w means that the two musical signals have similar complexity (or same local fluctuations) in the temporal scale. Hence, multifractal spectral width can be considered as one of the best parameter for the characterization of a music sample. In this paper we verify the presence of multifractality in the Tagore songs sung over five generations. We have taken 3 min clips of the complete rendition of 4 different Tagore songs which have been sung by artistes of different generations. Hence, we have divided the 3 minutes song signal into six equal segments and studied mutifractality using MFDFA technique in all the segments for all the signals. Each of the 30 sec segment was divided into 5 windows of 6 sec each and the average multifractal width has been given in **Table 2.**

**Table 2: Variation of Multifractal Width in the artistes of different generations**

|  | **Name of Artist** | **Part 1** | **Part 2** | **Part 3** | **Part 4** |
|---|---|---|---|---|---|
| Amaro Porano Jaha chay | Dinendra Thakur (1935) | 0.56 | 0.55 | 0.50 | 0.71 |
|  | Sagar Sen (1956) | 0.45 | 0.41 | 0.61 | 0.51 |
|  | Chinmoy Chatterjee (1973) | 0.51 | 0.36 | 0.64 | 0.84 |
|  | Suchitra Mitra (1986) | 0.46 | 0.30 | 0.39 | 0.20 |
|  | Jayati Chakraborty (2002) | 0.38 | 0.36 | 0.55 | 0.45 |
| Bahujuger Opar Hote | Bimalbhusan Mukherjee (1914) | 0.56 | 0.59 | 0.73 | 0.66 |
|  | Kanak Das (1945) | 0.68 | 0.52 | 0.56 | 0.55 |
|  | Debabrata Biswas (1963) | 0.52 | 0.32 | 0.47 | 0.42 |
|  | Mita Hoq (1977) | 0.46 | 0.39 | 0.40 | 0.33 |
|  | Rezwana Chowdhury (2008) | 0.42 | 0.78 | 0.80 | 0.68 |
| Krishnakoli Ami Tarei Boli | Shantidev Ghosh (1950) | 0.71 | 0.67 | 0.88 | 0.75 |
|  | Suchitra Mitra (1976) | 0.45 | 0.49 | 0.49 | 0.45 |
|  | Hemanta Mukherjee (1985) | 0.58 | 0.73 | 0.62 | 0.75 |
|  | Kamalini Mukherjee (1999) | 0.64 | 0.60 | 0.72 | 0.67 |
|  | Sasha Ghoshal (2010) | 0.68 | 0.58 | 0.73 | 0.66 |
| Maharaj Eki Saje | Amala Das | 0.68 | 0.43 | 0.43 | 0.48 |
|  | Subinoy Ray | 0.61 | 0.62 | 0.68 | 0.70 |
|  | Debabrata Biswas | 0.67 | 0.74 | 0.61 | 0.66 |
|  | Mohan Singh | 0.56 | 0.43 | 0.49 | 0.54 |
|  | Sahana Bajpayee | 0.53 | 0.57 | 0.67 | 0.48 |

The following figures show the variation of complexities for individual songs across five generations:

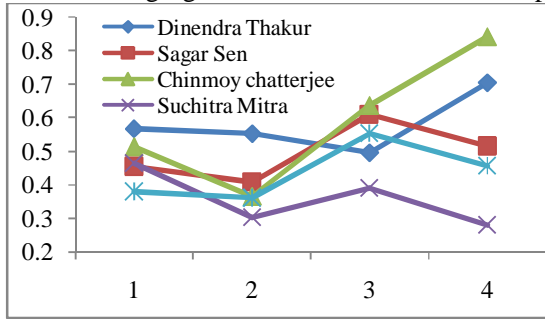

**Fig. 1a: Variation of complexities across different generations for the song *Amaro Porano Jaha Chay***

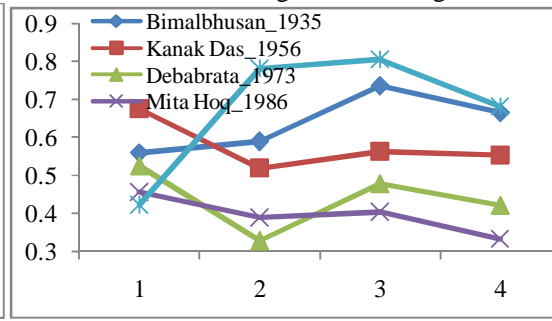

**Fig. 1b: Variation of complexities across different generations for the song *Bohujuger Opar Hote***

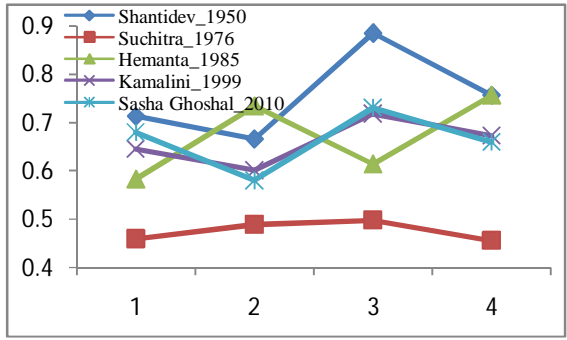
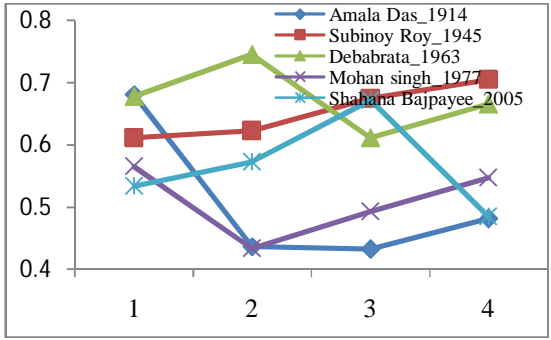

**Fig. 1c:** Variation of complexities across different generations for the song *Krishnakoli Ami Tarei Boli*

**Fig. 1d:** Variation of complexities across different generations for the song *Maharaj Eki saje*

From **Fig. 1a**, it is clear that there is significant variation of multifractal widths across different generation of singers in various parts. Initially, in the first two parts we see that the multifractal widths form a cluster, i.e. the singing styles of these singers in these parts are same, while in the last parts there is wide variation in the singing patterns. Thus, it can be said that the improvisational cues in singing are inherent in the last parts which shows wide variation in complexities. In **Fig. 1b**, we see that the singing styles of the 1st generation and the last generation are almost similar, while that in the 3rd and 4th generation, the complexity is quite on the lower side. The complexity of the 2nd generation lies in between the two. For this song, we find that there is considerable variation in all the parts belonging to each of the five generations. For the song *Krishnakoli,* we see that the singing style of the 2nd generation stands apart from the other generations as the multifractal width in this case is much lower than the others. In the last two generations, we see that the singing styles are almost similar in all the parts. We see that the multifractal widths of the 1st generation are quite high compared to the other generations, while the style of the 1st generation is almost similar to the 3rd generation in the 2nd and the last part. For **Fig. 1d**, there is considerable variation of complexities in all the parts, but there exists some clusters where the complexities are almost similar. We find that in 2nd, 3rd and the last generation, the complexities are similar to each other, same is the case for first and last generation in the 4th part. The 1st parts of all the five generations are almost similar to each other.

To further quantify the nature of evolution, we have averaged the values in each of the generations, and plotted the variation of complexities in each of the time window. For each song, the values of multifractal widths are plotted for all the generation of singers taken in a group and plotted in **Fig. 2**.

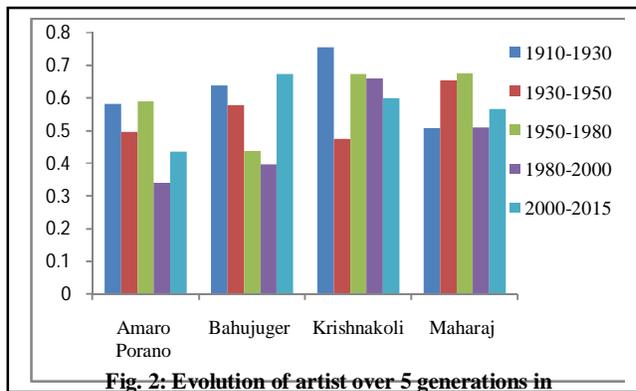

**Fig. 2:** Evolution of artist over 5 generations in each of the songs

From this figure, we see how the complexity in singing styles varies over different generations for the same song. Except for the last song, we find that the complexity values for the 1st generation is relatively high compared to the other generations. For the last song, we find that the complexity values of 1st and 4th generation are almost identical, while for the 1st song, the complexity values are identical for the 1st and the 3rd generation. This proves that the singing styles of these particular generations of singers are completely similar, while there is considerable variation among the other generations. It is also seen that the variation of multifractal widths is much less among the last generation of singers. For the songs *"Bahujuger Opar Hote"* and *"Krishnakoli."*, we see that the complexity values are almost similar for the 2nd and 3rd generation of singers. Thus, with the help of this technique we can have a quantified evolution of Tagore songs over a century in the form of complexity measures.

## CONCLUSION:

In this study, we provide a robust scientific technique with which we can have a quantitative measure of how the singing styles of different Tagore songs vary over the generations. The significant findings are:

1. It is interesting to see, that despite the presence of a robust notation system there is significant inter-artist variation even when the artistes of the same generation is singing the same song. Herein is hidden the self-improvisation of that particular artist where he/she uses his own skills to build something afresh from the omnipresent notation system.

2. In many cases, we find that the singing styles of the 1$^{st}$ generation and the last generation of singers are strangely similar; this may be caused due to the fact that singers of these two generations were not constrained to follow any stringent rules.

In conclusion, it can be said that the multifractal spectral width of each song can be developed as a parameter whose variation may give a cue about the singing style of that particular song/singer. Also, the change in the singing style of Tagore songs over a period of 100 years can be beautifully represented taking the help of complexity values corresponding to that song. Analysis of greater number of songs and artistes will lead to a more conclusive result.

## ACKNOWLEDGEMENT:

The first author, SS acknowledges the West Bengal State Council of Science and Technology (WBSCST), Govt. of West Bengal for providing the S.N. Bose Research Fellowship Award to pursue this research (193/WBSCST/F/0520/14). One of the authors, AB acknowledges the Department of Science and Technology (DST), Govt. of India for providing (A.20020/11/97-IFD) the DST Inspire Fellowship to pursue this research work.

## REFERENCES:

[1] Paul, S. K. "A Panoramic View of Rabindranath Tagore's Poems and His Poetic Sensibility." *Recritiquing Rabindranath Tagore* (2006): 1.

[2] Tagore, Rabindranath; Dutta, K. (editor); Robinson, A. (editor) (1997), *Rabindranath Tagore: An Anthology*, Saint Martin's Press (published November 1997), ISBN 978-0-312-16973-2

[3] Jun-ming, YUE Zhi-hua HUO. "Self-translation: a Special Form of Translation: Differentiation and Analysis on Tagore's Poetry Self-translation [J]." *Journal of Beijing Institute of Education* 2 (2008): 015.

[4] Ray, Sitansu. "Tagore, Freud and Jung on artistic creativity: A psycho-phenomenological study." *Life in the Glory of Its Radiating Manifestations*. Springer Netherlands, 1996. 329-341.

[5] Sengupta, Ranjan, et al. "On The Scientific Approach To Understand Improvisation: A Pilot Study."

[6] Bigerelle, Maxence, and A. Iost. "Fractal dimension and classification of music." *Chaos, Solitons & Fractals* 11.14 (2000): 2179-2192.

[7] Su, Zhi-Yuan, and Tzuyin Wu. "Music walk, fractal geometry in music." *Physica A: Statistical Mechanics and its Applications* 380 (2007): 418-428

[8] Peng, C-K., et al. "Mosaic organization of DNA nucleotides." *Physical Review E* 49.2 (1994): 1685.

[9] Kantelhardt, Jan W., et al. "Multifractal detrended fluctuation analysis of nonstationary time series." *Physica A: Statistical Mechanics and its Applications* 316.1 (2002): 87-114.